\documentclass[11pt]{article}
\usepackage{amsmath}
\usepackage{amssymb}
\usepackage{amsfonts}
\usepackage{latexsym}
\usepackage{graphicx}
\usepackage{enumerate}
\usepackage{multirow}
\usepackage{subfigure}

\newcommand{\R}{{\mathbb R}}

\newcommand{\Z}{{\mathbb Z}}
\newcommand{\N}{{\mathbb N}}
\newcommand{\C}{I}
\newcommand{\be}{\begin{equation}}
\newcommand{\ee}{\end{equation}}
\newcommand{\bes}{\begin{equation*}}
\newcommand{\ees}{\end{equation*}}
\newcommand{\lbl}{\label}
\newcommand{\ds}{\displaystyle}

\newtheorem{theorem}{Theorem}

\newtheorem{lemma}[theorem]{Lemma}

\renewcommand{\mod}{\text{ mod }}
\newcommand{\kh}{2 \pi \hat{K}^{(q)}(k)}

\renewcommand{\hat}{\widehat}

\def\qed{$\square$}
\def\one{\mathbf{1}}
\def\pf{{\it Proof.\;}}
\DeclareMathOperator{\re}{Re}

\begin{document}
\title{Stability of twisted states in the continuum Kuramoto model}
\author{Georgi S. Medvedev  and J. Douglas Wright  
\thanks{
Department of Mathematics, Drexel University, 3141 Chestnut Street,
Philadelphia, PA 19104;
{\tt jdoug@math.drexel.edu},
{\tt medvedev@drexel.edu}
}
}
\date{}
\maketitle
\begin{abstract}
We study a nonlocal diffusion equation approximating the dynamics of
coupled phase oscillators on large graphs. Under appropriate 
assumptions, the model has a family of steady state solutions called
twisted states. We prove a sufficient condition for stability of
twisted states with respect to perturbations in the Sobolev and BV
spaces. As an application, we study stability of twisted
states in the Kuramoto model on small-world graphs.
\end{abstract}
\section{Introduction}
\setcounter{equation}{0}

The Kuramoto model (KM) of coupled phase oscillators provides an
important framework for studying collective dynamics in a variety of systems 
across physics and biology \cite{Kur84, StrSync}.
To formulate the KM on a sequence of graphs, we first review the relevant 
background from the graph theory. Let $\Gamma_n=\langle V(\Gamma_n), E(\Gamma_n)\rangle$
be a graph on $n$ nodes, $V(\Gamma_n)=\{1,2,\dots,n\}=:[n]$. The pairs of connected 
nodes form the edge set of $\Gamma_n,$ $E(\Gamma_n)$. The KM on the following two
graph sequences will be used below to motivate the analysis and to illustrate the results.
The $k$--nearest-neighbor graph, $C_{n,k}$, $n\ge 2,$
$ 0\le k\le \lfloor n/2\rfloor,$ our first example, can be
geometrically  constructed 
by arranging $n$ nodes around a circle and connecting
each node to its $k$ neighbors from each side. Thus,  
$C_{n,k}=\langle [n], E(C_{n,k})\rangle$  such that for $1\le
i\neq j\le n$
$$
\{i,j\}\in E(C_{n,k}) \;\; \mbox{if}\;\; d_n(i,j):=\min\{|j-i|, 1-|i-j|\} \le k. 
$$

The small-world graph, $S_{n,k,p}$ with $n$ and $k$ as above and $p\in [0,1]$
is obtained from $C_{n,k}$ by replacing each edge with probability $p$ (independently
from other edges) by a random edge \cite{WatStr98}.
There are several variants of small-world graphs, which differ in
technical details of how the random edges are selected (see \cite{WatStr98,NewWat99}).
In this paper, we follow \cite{Med14b}, where small-world graphs are
interpreted as W-random graphs. Specifically, for a given $n\gg 1,$
$k\in [\lfloor n/2\rfloor]$ and $p\in (0,1]$, $S_{n,k,p}=\langle [n],
E(S_{n,k,p})\rangle$ such that
$$
\operatorname{Prob}(\{i,j\}\in E(S_{n,k,p}))=\left\{ \begin{array}{ll} 1-p, & d(i,j)\le
    k,\\
p,& \mbox{otherwise}.
\end{array}
\right.
$$

Let us scale the number of neighbors 
$ k=\lfloor rn\rfloor$ for $r\in (0,1/2]$. The resultant  graph sequences 
$\{C_{n,\lfloor rn\rfloor}\}$ and $\{S_{n,\lfloor rn\rfloor,p}\}$  have well-defined asymptotic
behavior in the limit as $n\to\infty$. In fact, $\{C_{n,\lfloor rn\rfloor}\}$ and 
$\{S_{n,\lfloor rn\rfloor,p}\}$ are convergent sequences of dense graphs \cite{LovGraphLim12}.  
The asymptotic properties of convergent graph sequences are captured by a symmetric
function on a unit square called a graphon. The graphons corresponding
to the graph limits of $\{C_{n,\lfloor rn\rfloor}\}$ and
$\{S_{n,\lfloor rn\rfloor,p}\}$ are piecewise constant functions, which are explained in 
Fig.~\ref{f.1}\textbf{a}.

\begin{figure}
\begin{center}
\textbf{a}\hspace{0.2 cm}\includegraphics[height=2.0in,width=1.8in, angle=90]{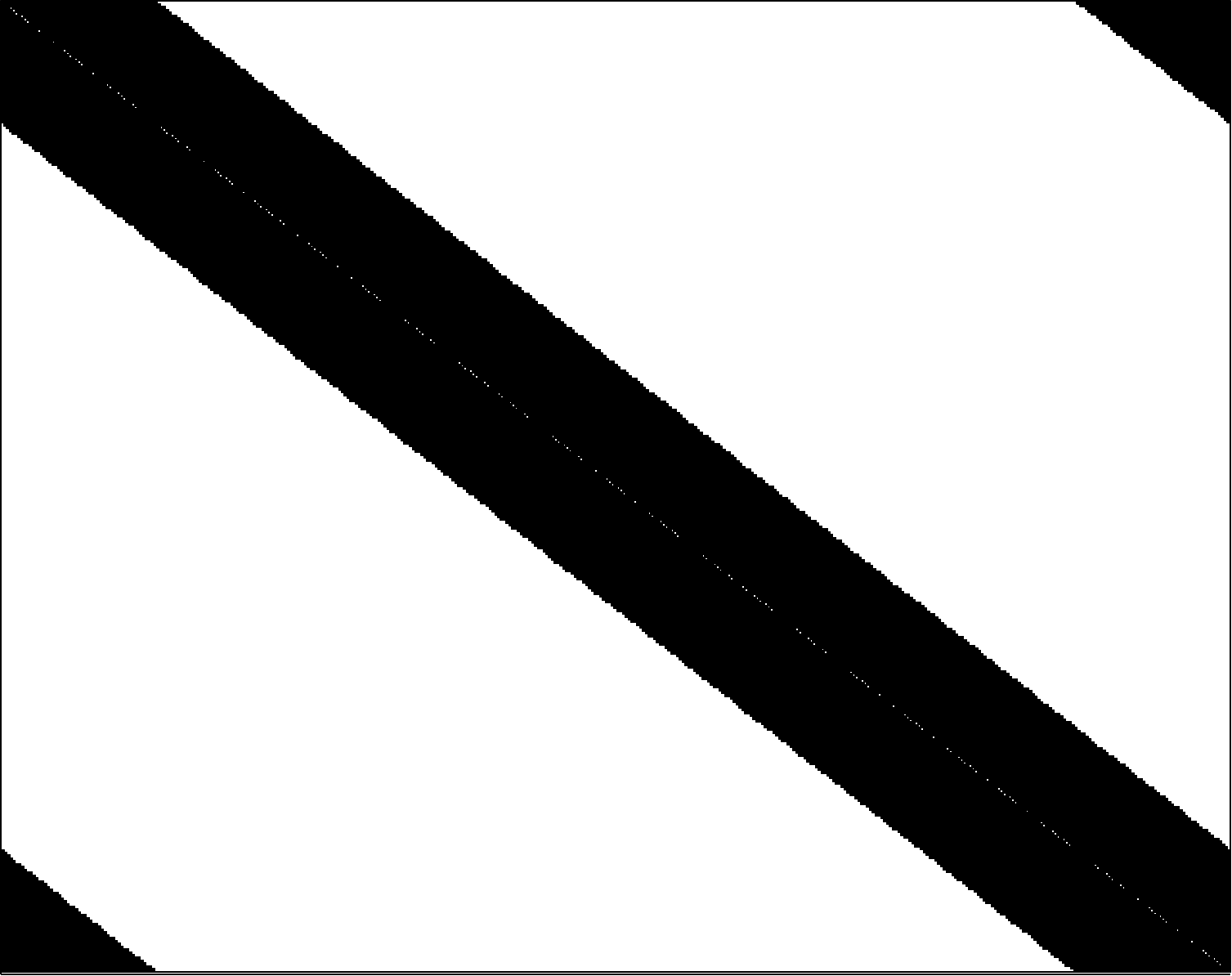}
\hspace{0.5 cm}
\textbf{b}\hspace{0.1 cm}\includegraphics[height=1.8 in, width=2.0in]{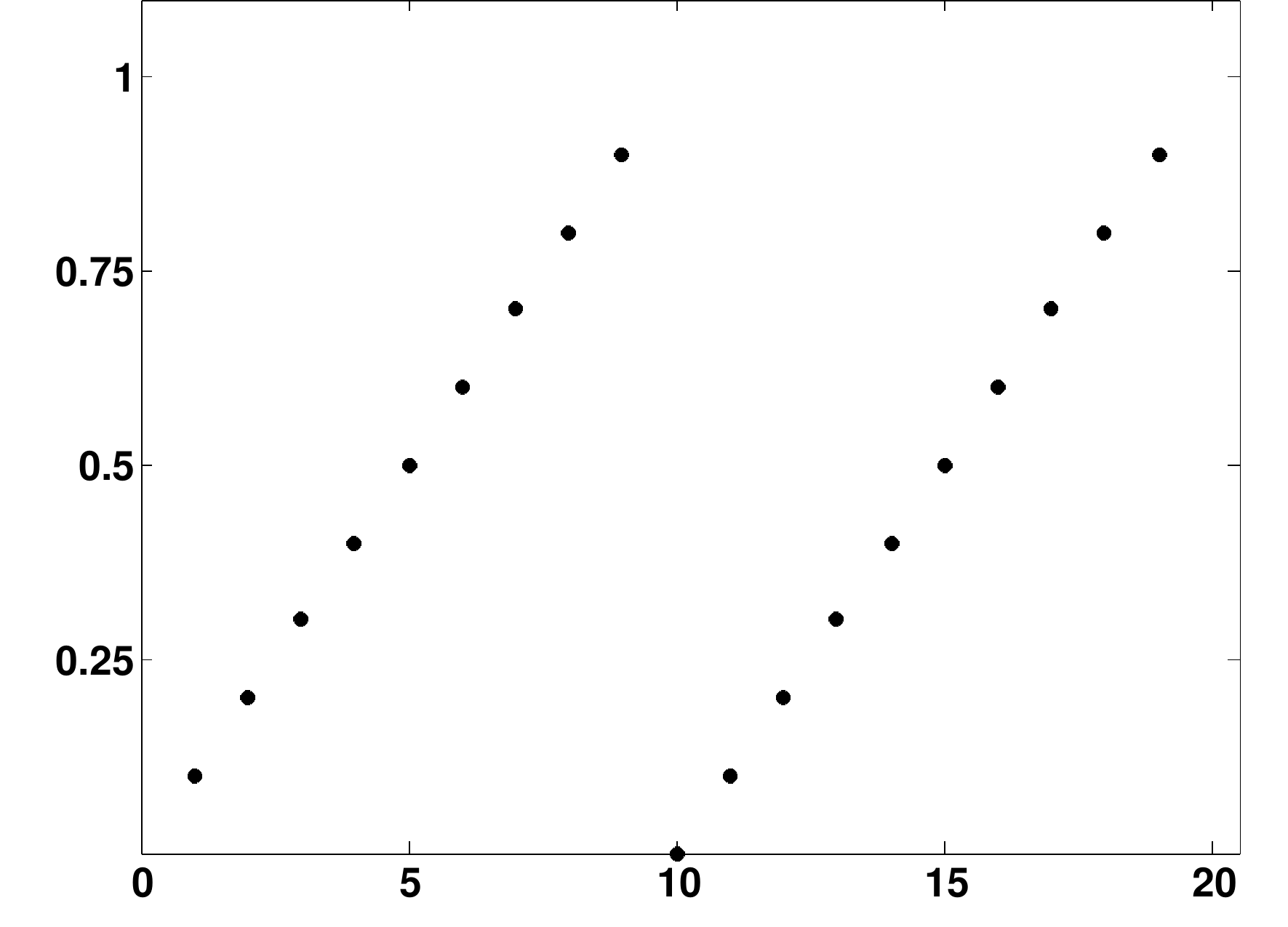}
\end{center}
\caption{ \textbf{a)} The limiting graphon  for $\{C_{n,\lfloor rn\rfloor}\}$ is a $\{0,1\}$--valued function
on the unit square, whose support is shown in black. The limiting graphon of 
$\{S_{n,\lfloor rn\rfloor,p}\}$ is equal to $1-p$ over the black region and  $p$ otherwise.
\textbf{b)} A $2$--twisted state for a KM with twenty oscillators.
}
\lbl{f.1}
\end{figure}

The KM of coupled identical phase oscillators on the graph sequence $\{\Gamma_n\}$ 
is defined as follows
\be\lbl{omega-Kuramoto}
\dot u^{\omega}_{ni}=\omega + {1\over n} \sum_{j:\{i,j\}\in E(\Gamma_n)} \sin
(2\pi( u^\omega_{nj}- u^\omega_{ni})),\; i\in [n],
\ee
where $u_{ni}$ and $\omega$ stand for the phase  and the intrinsic frequency of the oscillator $i$ 
respectively. By switching to the rotating frame of coordinates, we  eliminate $\omega$ in
\eqref{omega-Kuramoto}. Thus,
\be\lbl{discrete-Kuramoto}
\dot u_{ni}= {1\over n} \sum_{j:\{i,j\}\in E(\Gamma_n)} \sin
(2\pi(u_{nj}-u_{ni})),\; i\in [n].
\ee
If $\Gamma_n$ is a Cayley graph on a cyclic group,
\eqref{discrete-Kuramoto} has a family 
of steady-state 
solutions 
\be\lbl{q-twist}
u_{ni}^{(q,c)}= \left( {qi\over n}+c\right) \mod 1,\;\; i\in [n],
\ee
where $q$ is an integer between $-n+1$ and $n-1$ and $c\in \R$ \cite{MedTan15b} (see
Fig.~\ref{f.1}\textbf{b}).
If $q=0$ then  $u^{(0,c)}_{n}:=(c,c,\dots,c)$  is a spatially homogeneous solution. For $q\neq 0$,
$u_{ni}^{(q,c)}, i\in [n]$ wind around the circle $|q|$ times. In the original frame of coordinates,
$u^\omega_{ni}=u_{ni}^{(q,c)}+\omega t$ form uniformly twisted
travelling waves. Hence, $u^{(q,c)}_n$  are
called twisted states. It was pointed out in \cite{WilStr06} that the stability of twisted states
depends on the connectivity of $\{\Gamma_n\}$. For instance, for the KM on $\{C_{n,k}\}$
the stability depends on the number of neighbors $k$.  Similarly, 
stability of the twisted states can be linked to the network topology
in the KM on many other 
families of graphs including Cayley, Erd\H{o}s--R{\' e}nyi, and small--world graphs \cite{Med14c, MedTan15b}.

The KM on convergent graph sequences like $\{C_{n,\lfloor nr\rfloor }\}$ and 
$\{S_{n, \lfloor nr\rfloor,p}\}$, for large $n$ can be approximated  by the nonlocal diffusion equation
\be\label{Kur}
\partial_t u(x,t) = \int_I W(x,y) \sin(2\pi(u(y,t)-u(x,t))) dy, \;\; I:=[0,1],
\ee
where $u(x,t)$ now describes the phases of the continuum of oscillators distributed over $I.$
The constructions of the continuum limits for the KM on $\{C_{n,k}\}$
and $\{S_{n,k,p}\}$ are explained in \cite{Med14a, Med14b,
  Med14c}. Here, we go over a few details that are relevant to the
stability problem considered in this work.
The kernel $W: I^2\to \R$ in the integral on the right-hand side of
\eqref{Kur} is the graphon describing the limit of the underlying graph 
sequence. 

For $\{C_{n,\lfloor nr\rfloor }\}$ or 
$\{S_{n, \lfloor nr\rfloor,p}\},$ the limiting graphon is given by 
$
W(x,y) = K(y-x),
$
where $K: \R \to \R$ is a $1-$periodic function.
Specifically,  for $\{S_{n, \lfloor nr\rfloor,p}\}$, we have $ W(x,y)= K_{r,p}(y-x)$,
with $K_{r,p}$ defined on the interval $(-1/2, 1/2]$ by
\be\lbl{Kr}
K_{r,p}(x) = \left\{ \begin{array}{ll} 1-p, & \mbox{if}\; |x|\le r,\\
                                              p, &  \mbox{otherwise},
\end{array}
\right.
\ee
and extended to $\R$ by periodicity. For  $\{C_{n,\lfloor nr\rfloor }\}$ we have 
$W(x,y) =K_r(y-x):=K_{r,0}(y-x).$ 
Thus, the continuum limit for the KM on these two graph sequences can
be rewritten as
\be\lbl{KurS}
\partial_t u(x,t) = \int_{I} K(y-x) \sin(2\pi(u(y,t)-u(x,t))) dy
\ee
where $K:\R \to \R$ is a given $1-$periodic function. Likewise 
$u(x,t)$ is $1$-periodic with respect to $x$. 

If $K$ is an even function, as in the case of $K=K_{r,p}$, the
continuum model \eqref{KurS} has a family of steady-state solutions 
\be\lbl{def-q}
u^{(q,c)}(x):=(qx + c) \mod 1, \; q\in \Z, c\in \R,
\ee
i.e., the continuous twisted states; we henceforth assume that $K$ is even.

We are interested in stability of twisted states
\eqref{def-q}. The following sufficient condition for the linear stability 
of the continuous twisted states was derived in \cite{WilStr06}: 
\be\lbl{linear}
\lambda(q,m):=\tilde K(m+q)-2\tilde K(q)+\tilde K(q-m) \le \lambda_q<0,
\;\quad\exists \lambda_q,\ \forall m\in \N,
\ee
where $\tilde K(m)=\int_I K(x) \cos(2\pi m x) dx$. 
In this work, we prove that \eqref{linear}, in fact, implies nonlinear stability 
 in $H^1_{per} $ (the usual Sobolev space of 
once weakly differentiable $1-$periodic functions)
or $BV_{per}$ (the space of $1-$periodic functions of bounded variation).
Specifically we have the following theorem.

\begin{theorem} \lbl{thm.stab}
Suppose that $K$ is an even function in $L^2_{per}$.
Let $X =  H^1_{per}$ or $BV_{per}$.
Suppose that condition \eqref{linear} holds.  Then there exists
$\delta>0$, $b>0$ and $C>0$ such that $\| u_0 - \mu - u^{(q,c)}\|_{X} \le \delta$ implies 
$$
\sup_{t \ge 0} e^{bt} \|u(\cdot,t) - \mu - u^{(q,c)}(\cdot)\|_{X} \le C\| u_0 - \mu -u^{(q,c)}\|_{X}.
$$
Here $\mu := \ds\int_I (u_0(y) - u^{(q,c)}(y)) dy$ and $u(x,t)$ is the solution of \eqref{KurS} with $u(x,0) = u_0(x)$.
\end{theorem}

The proof of Theorem~\ref{thm.stab} 
is based on a general stability
theorem  which we state and prove 
in Section~\ref{sec.nonlinear}  (specifically, see Theorem \ref{absthm} below).  
This general stability theorem is a more or less classical ``linear implies nonlinear stability" theorem
of the sort used to prove asymptotic stability of fixed points in ordinary differential equations.
The main technical difficulty in this work is showing that 
\eqref{KurS} and its $q-$twisted states satisfy the hypotheses of
 the general stability theorem.

The nonlocal equation \eqref{KurS} does not possess the strong
smoothening property, which facilitates stability analysis of spatial structures in 
parabolic equations, a closely related class of models \cite{Henry}.
Solutions of the initial value problem for \eqref{KurS} may  have poor spatial
regularity if the kernel $K$ and the initial data are not smooth enough (cf.~\cite[Theorem~3.3]{Med14a}).
Therefore, along with stability with respect to sufficiently regular perturbations from  $H^1_{per}$,
we feel it is important to understand the stability of $q-$twisted states with respect to rough
perturbations. To this end, in the second part of the paper, we study stability of the $q-$twisted states
to perturbations from $BV_{per}$. 
The Sobolev inequalities and the natural connection of $H^1_{per}$ to Fourier series make much of our stability analysis in $H^1_{per}$  
pretty straightforward; after all, the linear stability condition \eqref{linear} is  a condition on the Fourier coefficients of $K$.
On the other hand, Fourier analysis for $BV_{per}$ functions is not as simple as it is in $H^1_{per}$ and this carries with it several challenges, of which the
main one is showing that the linearization of \eqref{KurS} about a $q-$twisted states generates an appropriately contractive semigroup.
Addressing all of these technical considerations takes place in Section~\ref{sec.stab}, which contains the proof of Theorem~\ref{thm.stab}.

We
conclude this paper with the application to stability of twisted
states in the continuum KM on small-world graphs in Section~\ref{sec.examples}.

\section{Linear implies nonlinear stability}
\lbl{sec.nonlinear}
\setcounter{equation}{0}

Theorem~\ref{thm.stab} is a  consequence of the following abstract stability result:
\begin{theorem} \label{absthm} 
Suppose that $X$ is a Banach space. Suppose there exist $C_1,C_2>0$, $\rho>0$ and $b>0$ such the following hold.
\begin{enumerate}[(i)]
\item $L$ is a bounded linear map from $X$ to itself.
\item The uniformly continuous semigroup generated by $L$, given by $e^{Lt}$, has, for all $t > 0$, $\| e^{Lt}\|_{X \to X} \le C_1 e^{-bt}$.
\item $N$ is a continuous map from the ball of radius $\rho$ in $X$, denoted $B_\rho(0)$, back into $X$.
\item $N(0) = 0$.
\item For all $f,g  \in B_\rho (0)$ we have $\|N(f) - N(g)\|_{X} \le C_2\left(\|f\|_X + \|g\|_X \right) \| f- g \|_X$.
\end{enumerate}
Then there exists  $\delta \in  (0,\rho]$ and $C_3 > 0$ such that the following hold for all $\xi_0 \in B_\delta(0) \subset X$.
\begin{enumerate}[(a)]
\item There exists a unique $\xi \in C^1(\R^+;X)$ for which $\xi(0) = \xi_0$ and 
$\ds
\partial_t \xi = L \xi + N(\xi)
$
for all $t > 0$.
\item 
$\ds
\sup_{t \ge 0} e^{bt} \|\xi(t)\|_{X} \le C_3 \| \xi_0\|_X.
$
\end{enumerate}

\end{theorem}
The proof is classical (see, for instance, \cite{Henry}), but included for completeness.  
 
\noindent{\it Proof of Theorem \ref{absthm}:}
 Let 
 $$
 Y:= \left\{ f \in C(\R^+;X) : \sup_{t \ge 0} e^{bt} \|f(t)\|_X =: \| f\|_{Y} < \infty\right\}.
 $$
Note that $Y$, equipped with $\| \cdot \|_Y$, is a Banach space.
Let
$$\ds
\gamma:=\max\left\{{b \over 4 C_1C_2},1,\rho \right\}\quad \text{and} \quad \delta := \max\left\{{3\gamma \over 4 C_1},\rho \right\}
$$
and fix $\xi_0 \in X$ with $\| \xi_0 \|_X < \delta.$ Finally put
$r:=\ds {4\over 3} C_1 \|\xi_0\|_X$. Note that $r<\gamma$. 

For $f \in Y$ with $\|f\|_Y < \gamma$ define:
 $$
 \Psi(f):= e^{Lt} \xi_0 + \int_0^t e^{L(t-s)} N(f(s))ds.
 $$
We claim that $\Psi$ maps $B_r:=\left\{f \in Y : \|f\|_Y < r \right\}$ back into itself and satisfies a contraction estimate on this set.
If the claim is true then Banach's fixed point theorem implies that $\Psi$ has a unique fixed point in $B_r$ which we denote $\xi$.
Differentiation of $\xi = \Psi(\xi)$ with respect to $t$ shows that $\partial_t \xi = L \xi + N(\xi)$. This also shows that $\xi \in C^1(\R^+;X)$.
Moreover $\xi(0) = \Psi(\xi) \big \vert_{t = 0} = \xi_0.$ Thus we have the conclusion (a). Membership of $\xi \in Y$ implies the conclusion (b).

Thus we need only establish the claim. Assume that $f,g \in B_r$.
The semigroup estimate (ii) and estimate (v) for $N$ give:
 \begin{equation}\label{contract1}
 \begin{split}
 \|\Psi(f) - \Psi(g)\|_Y\le &\sup_{t \ge 0} e^{bt}\left\| \int_0^t  e^{L(t-s)} [N(f(s)) - N(g(s))]  ds\right\|_X\\
 \le &C_1 \sup_{t \ge 0} e^{bt} \int_0^t  e^{-b(t-s)} \| N(f(s)) - N(g(s))\|_X ds\\
 \le & C_1C_2\sup_{t \ge 0} \int_0^t  e^{bs} \left( \|f(s)\|_X + \|g(s)\|_X\right) \|f (s) - g(s)\|_X ds\\
  \le & C_1C_2\sup_{t \ge 0} \int_0^t  e^{-bs} \left( \|f\|_Y + \|g\|_Y\right) \|f  - g\|_Y ds\\
  =& b^{-1} C_1C_2\left( \|f\|_Y + \|g\|_Y\right) \|f  - g\|_Y.
\end{split}
\end{equation}
Since $f,g \in B_r$ and $r < \gamma$ this implies
$\ds
 \|\Psi(f) - \Psi(g)\|_Y\le2b^{-1} C_1C_2 \gamma \|f-g\|_Y.
$
The definition of $\gamma$ above then gives:
\begin{equation}\label{contract2}
\|\Psi(f) - \Psi(g)\|_Y \le {1 \over 2} \|f-g\|_Y.
\end{equation}
Thus $\Psi$ satisfies a contraction estimate on $B_r$. 

To show that $\Psi$ maps $B_r$ into itself, we first
use the semigroup property (ii) and the fact that $N(0) = 0$ to get:
 \begin{equation}\label{zero goes}
\| \Psi(0)\|_Y = \sup_{t \ge 0} e^{bt} \| e^{Lt} \xi_0\|_X \le C_1 \|\xi_0\|_X. 
 \end{equation}
Then we use \eqref{contract1} in combination with this to get:
\be\label{toself}
\|\Psi(f)\|_Y \le \|\Psi(f) - \Psi(0)\|_Y + \|\Psi(0)\| \le b^{-1}C_1C_2 \| f \|_Y^2 +  C_1 \|\xi_0\|_X
\ee
Since $\|f \|_Y < r < \gamma \le b/4C_1C_2$ we have: 
$\ds
\|\Psi(f)\|_Y < {1 \over 4} r +  C_1 \|\xi_0\|_X.
$
Then we use the defintion of $r$ to conclude that
$\ds
\|\Psi(f)\|_Y < {1 \over 4} r +  {3 \over 4} r = r. 
$
Thus $\Psi$ maps $B_r$ to $B_r$ and we are done.

\section{Proof of Theorem~\ref{thm.stab}}
\lbl{sec.stab}
\setcounter{equation}{0}

Fix $q \in \Z^+$ and ${c \in \R}$. Suppose that $u(x,t) = u^{(q,{c})}(x) + {\eta}(x,t)$ where $u(x,t)$ solves \eqref{KurS}
and $\eta(x,t)$ is periodic (with period $1$) in $x$.
A routine computation show that $\eta(x,t)$ solves:
\be\label{eta eqn v1}
\partial_t \eta = \Phi(\eta)
\quad \text{where} \quad
\Phi(\eta)(x):=\int_I K(y-x) \sin( 2 \pi [q(y-x) + \eta(y) - \eta(x) ]) dy.
\ee
The following technical lemma is proved in Section \ref{Phi props}:
\begin{lemma}\label{Phi}
Let $K \in L^2_{per}$. Let $X$ be either $BV_{per}$ or $H^1_{per}$. Then the map
$$
\Phi(\eta)(x):=\int_I K(y-x) \sin( 2 \pi [q(y-x) + \eta(y) - \eta(x) ]) dy
$$
is $C^\infty$ from $X$ to itself.  Moreover each of its derivatives is uniformly Lipschitz on any bounded subset of $X$.
Lastly
$
\Phi'(0) =: L
$
where
$$
L\eta(x) :=2 \pi \int_I K\left(y-x\right) \cos\left(2 \pi q\left(y-x\right)\right) \left(\eta\left(y\right) - \eta\left(x\right)\right)dy. 
$$
\end{lemma}
Note that this lemma implies that the initial value problem for \eqref{eta eqn v1} is well-posed in $BV_{per}$ and $H^1_{per}$.

Next we have
\begin{lemma}\label{the goods are odd}
Let $K \in L^2_{per}$ be an even function. 
Then for any $\eta \in BV_{per}$ or $H^1_{per}$ we have
$$
\int_I \Phi(\eta)(x) dx = \int_I L\eta(x) = 0.
$$
\end{lemma}

\noindent{\it Proof of Lemma \ref{the goods are odd}:} 
We prove the result only for $\Phi$ as the result for $L$ goes along the same lines.
It is clear that
\be\begin{split}
\int_I \Phi(\eta)(x) dx &= \int_I  \int_I K(y-x) \sin( 2 \pi [q(y-x) + \eta(y) - \eta(x) ]) dy dx \\
&=  \int_I  \int_I K(x-y) \sin( 2 \pi [q(x-y) + \eta(x) - \eta(y) ]) dx dy.
\end{split}
\ee
Since $K$ is even and sine is odd, applying Fubini's theorem to the above gives:
\be\begin{split}
 \int_I \Phi(\eta)(x) dx &=-  \int_I  \int_I K(y-x) \sin( 2 \pi [q(y-x) + \eta(y) - \eta(x) ]) dy dx= - \int_I \Phi(\eta)(x) dx.
\end{split}
\ee
This implies that $\ds \int_I \Phi(\eta)(x) dx = 0$. \hfill $\square$
%
%
%

This last lemma  implies that a solution $\eta(x,t)$ of \eqref{eta eqn v1} meets 
$$
\int_I {\eta}(x,t) dx = \int_I {\eta}(x,0) dx=:\mu
$$
for all $t$. 
Now set
$$
\xi(x,t) := \eta(x,t) - \mu.
$$
Clearly $\ds \int_I \xi(x,t) dx = 0$ for all $t$. Also observe that
$
\partial_t \xi = \Phi(\xi)
$
since $\partial_t \xi = \partial_t \eta$ and $\xi(x,t)-\xi(y,t) = \eta(x,t)-\eta(y,t)$. 

If we put 
$$
N(\xi):=\Phi(\xi) - L\xi
$$
then
clearly
\be\label{xi eqn}
\partial_t \xi = L \xi + N(\xi).
\ee
If we can establish properties (i)-(v) in Theorem \ref{absthm} for $L$ and $N$ as defined above, then the conclusions of that theorem immediately imply our main result, Theorem \ref{thm.stab}.
To be clear, we will establish these properties on the spaces
$$
H^1_{per,0}:=\left\{f \in H^1_{per} : \int_I f(x) dx = 0 \right\}
$$
and
$$
BV_{per,0}:=\left\{f \in BV_{per} : \int_I f(x) dx = 0  \right\}.
$$
These are closed subspaces of Banach spaces and thus are themselves Banach spaces with the appropriate inherited norm. Lemma \ref{the goods are odd} implies that $L$ and $N$
 are well defined maps on these spaces. Thus \eqref{xi eqn} is well-posed on either of these spaces.

\subsection{Estimates for $N$}
The estimates for $N$ are relatively simple, given Lemma \ref{Phi}. In particular, that lemma immediately implies property (iii) of Theorem \ref{absthm} holds for any $\rho >0$.
Moreover, by construction
$
N'(0) = \Phi'(0) - L = 0,
$
which is to say that property (iv) is satisfied.
Property (v) follows from an invocation of the ``difference of squares" estimate:
\begin{lemma}\label{diff of sqs}
Suppose that $X$ is a Banach space and $U \in X$ is open, convex and contains 0. Suppose that $N:U \to X$ is $C^{1,1}$---that is to say, $N'$ is uniformly Lipschitz continuous on $U$.
Furthermore suppose that $N'(0) = 0$. Then there exists $C>0$ such that for all $f,g \in U$ we have
$$
\| N(f) - N(g)\|_{X} \le C\left( \| f\|_X + \|g\|_X\right) \left\| f- g \right \|_{X}.
$$
\end{lemma}
\noindent{\it Proof of Lemma \ref{diff of sqs}:} 
Fix $f,g \in U$. Since $N$ is differentiable we have by the fundamental theorem of calculus and chain rule:
$$
N(f) - N(g) = \int_0^1 {d \over dt}\left( N (g + t(f-g)) \right) dt =  \int_0^1 N' (g + t(f-g))\left(f-g \right) dt.
$$
Note that convexity of $U$ implies $g + t(f-g)\in U$ for all $t\in [0,1]$.
Since $N'(0) =0$ we have
$$
N(f) - N(g)  =  \int_0^1\left[ N' (g + t(f-g))-N'(0)\right]\left(f-g \right) dt.
$$
Since $N'$ is uniformly Lipschitz on $U$ (with constant $C_L\ge 0$, say), we have by the triangle inequality:
$$
\| N(f) - N(g)\|_X  \le C_L \int_0^1\| g + t(f-g)\|_{X} \| f-g\|_{X}dt \le C\left( \| f\|_X + \|g\|_X\right) \left\| f- g \right \|_{X}.
$$
That completes the proof.
\hfill $\square$

Thus we have established properties (iii)-(v) for $N$ in Theorem \ref{absthm}. 

\subsection{Estimates for $L$}
Property (i) in Theorem \ref{absthm} is implied by Lemma \ref{Phi}.
Next observe that
$$
L \xi = K^{(q)} * \xi - \gamma \eta
$$
where the ``$*$" denotes the usual periodic convolution ({\it i.e.} if $f$ and $g$ are $1-$periodic then $f * g(x):=\int_I f(y-x) g(y) dy$). Above
$$
K^{(q)}(x) := 2\pi K(x) \cos(2 \pi q x)
\quad \text{and}\quad
\gamma := 2 \pi \int_I K(y) \cos(2 \pi y) dy.
$$

We will get a formula for the semigroup $e^{Lt}$ by means of the Fourier series.
For periodic functions $f(x)$, let
$$
\hat{f}(k):=\int_I f(x) e^{2\pi i k x} dx
$$
be the coefficients of its Fourier series. The Fourier inversion theorem
implies
$$
f(x) = \sum_k  \hat{f}(k)e^{-2 \pi i k x}
$$
where the equality is in the sense of $L^2_{per}$.

The convolution theorem gives
$$
\widehat{(L f)}(k) =\lambda(k) \hat{f}(k)
$$
where
$$
\lambda(k):= 2 \pi \widehat{K}^{(q)}(k) - \gamma.
$$
Thus we can define the semigroup $e^{Lt}$ by
\be\label{sg def}
e^{Lt} f (x):= \sum_{k\ne0} e^{ \lambda(k) t-2\pi ikx}  \hat{f}(k).
\ee
That is to say
$$
\hat{e^{Lt} f}(k) = e^{\lambda(k) t} \hat{f}(k).
$$

Condition \eqref{linear} implies that there exists $b \in \R$ such that
\be\label{stab cond}
\re \lambda(k) = \re (2 \pi \widehat{K}^{(q)}(k) - \gamma) \le -b < 0 
\ee
for all $k \ne 0$. We are considering mean zero functions and so the $k = 0$ mode is excluded, in any case.
Which is to say that the spectrum of $L$ is in the left half plane and thus the system is spectrally stable. 

\subsubsection{Semigroup estimates in $H^1_{per,0}$}
The semigroup estimates here are simple to establish, as the semigroup is a Fourier multiplier. By Plancheral's theorem we have, for $f \in H^1_{per,0}$,
$$
\| e^{Lt} f\|_{H^1_{per}}^2 = \sum_{k \ne 0} (1+k^2) \left \vert e^{ \lambda(k) t} \right \vert^2 \left \vert \hat{f}(k) \right\vert^2= \sum_{k \ne 0} (1+k^2) e^{ 2\re\lambda(k) t} \ \left \vert \hat{f}(k) \right\vert^2. 
$$
Using \eqref{stab cond} gives, for $t > 0$:
$$
\| e^{Lt} f\|_{H^1_{per}}^2 \le \sum_{k \ne 0} (1+k^2)  e^{-2 bt}  \left \vert \hat{f}(k) \right\vert^2 = e^{-2bt} \|f\|_{H^1_{per}}^2.
$$
Thus we have property (ii) for $X=H^1_{per,0}$. Given the validity of Lemma \ref{Phi}, this concludes the proof of Theorem \ref{thm.stab} for $X=H^1_{per,0}$.

\subsubsection{Semigroup estimates in $BV_{per,0}$}
Suppose that $f \in BV_{per}$. Recall that
$$
\|f\|_{BV_{per}}:=\int_I |f(x)| dx + V(f)
$$
where
$$
V(f):= \sup_{P\in\mathcal{P}(I)}  \sum_P |f(x_i) - f(x_{i-1})|
$$
and $\mathcal{P}(I)$ is the set of ordered partitions of $I$.
It is not true that Fourier multipliers with bounded symbols define bounded maps 
on $BV_{per}$.
But $e^{Lt}$ does, as the following computations show.

First observe that
$$
e^{\lambda(k) t} = e^{-\gamma t} + e^{-b t} \left( e^{(2 \pi \hat{K}^{(q)}(k) - \tilde{\gamma})t} - e^{-\tilde{\gamma} t} \right).
$$
In the above $\tilde{\gamma} := \gamma - b\ge0$. Note that \eqref{stab cond} implies that
$\re(2 \pi \hat{K}^{(q)}(k)- \tilde{\gamma})\le 0$ for all $k$.
The fundamental theorem of calculus gives:
$$
\hat{M}(k,t):=e^{(\kh - \tilde{\gamma})t} - e^{-\tilde{\gamma} t} = -t\int_{\tilde{\gamma}}^{\tilde{\gamma}-\kh} e^{-st} ds.
$$
We assume $t > 0$.
Note that since $\hat{K}^{(q)}$ may be complex-valued, the integral above is computed along the line segment in $\mathbb{C}$ connecting
$\tilde{\gamma}$ to $\tilde{\gamma}-\kh$ which is in the right half plane.
The ``$ML$-inequality"  gives:
$$
\left \vert 
\hat{M}(k,t)
\right \vert
\le |t|  \left \vert \kh  \right \vert.
$$
We have  $K \in L^2_{per}$ and thus so is $K^{(q)}$. Therefore
letting
$$
M(x,t):=\sum_k \hat{M}(k,t) e^{-2 \pi i k x}
$$
defines an $L^2_{per}$ function by virtue of the preceding estimate. Of course $\| M(t)\|_{L^2_{per}} \le C|t|\|K^{(q)}\|_{L^2_{per}}$.

Since $e^{Lt}$ is defined by \eqref{sg def}, the decomposition $e^{\lambda(k) t} = e^{-\gamma t} + e^{-bt} \hat{M}(k,t)$ gives, via the convolution theorem,
the following formula for the semigroup:
$$
e^{Lt} f = e^{-\gamma t} f + e^{-bt}M(\cdot,t)* f.
$$
Now fix $b' \in (0,b)$.  We have
$$
\| e^{Lt} f\|_{BV_{per}} \le e^{-\gamma t} \|f\|_{BV_{per}} + e^{-bt} \|M(\cdot,t)\|_{L^2_{per}} \|f\|_{BV_{per}} \le C_{b'} e^{-b't}\|f\|_{BV_{per}}.
$$
Thus we have property (ii) for $X=BV_{per,0}$. Given the validity of Lemma \ref{Phi}, this concludes the proof of Theorem \ref{thm.stab} for $X = BV_{per,0}$.

\subsection{Properties of $\Phi$}\label{Phi props}

This section will ultimately establish Lemma \ref{Phi}.
First we have:
\begin{lemma}\label{substitution operators}
Let $X =H^1_{per}$ or $BV_{per}$. Then
the maps
$$
\Xi(\eta)(x) = \cos(2 \pi \eta(x)) \quad \text{and} \quad S(\eta)(x) = \sin(2 \pi \eta(x))
$$
are $C^\infty$ from $X$ to itself. Moreover each derivative is uniformly Lipschitz on any bounded subset of $X$.
\end{lemma}

\noindent{\it Proof of Lemma \ref{substitution operators}:}
When $X=H^1_{per}$,  $S$ and $\Xi$ are Nemitskii substitution operators and the result is classical \cite{Renardy}. And so we only prove this result for $X = BV_{per}$. 

First note that
$$
\| S(\eta) \|_{BV_{per}} = \int_{\C}|\sin(\eta(x))|dx +  \sup_{P \in \mathcal{P}} \sum_P |\sin(\eta(x_i)) - \sin(\eta(x_{i-1}))|
$$
Since $|\sin(\theta)-\sin(\phi)| \le2 | \theta - \phi|$ for any $\theta,\phi$, the above gives:
$$
\| S(\eta) \|_{BV_{per}} \le C \int_{\C}|\eta(x)|dx +  C\sup_{P \in \mathcal{P}} \sum_P |\eta(x_i) - \eta(x_{i-1})| = C \| \eta \|_{BV_{per}}.
$$
Likewise
$$
\| \Xi(\eta) \|_{BV_{per}} \le C + C \| \eta \|_{BV_{per}}.
$$

Using the trigonometry identities $\sin(\theta) - \sin(\phi) = 2 \sin((\theta-\phi)/2) \cos((\theta+\phi)/2)$
and $\cos(\theta) - \cos(\phi) = -2 \sin((\theta-\phi)/2) \sin((\theta+\phi)/2)$, together with the algebra estimate $\| f g \|_{BV_{per}} \le C \| f\|_{BV_{per}} \| g\|_{BV_{per}}$,
we get
$$
\| S(\eta) - S(\xi) \|_{BV_{per}}  + \| \Xi(\eta) - \Xi(\xi) \|_{BV_{per}}  \le C\left(1+\|\eta\|_{BV_{per}} + \|\xi\|_{BV_{per}} \right) \| \eta - \xi\|_{BV_{per}}.
$$
Which is to say that both $S$ and $\Xi$ are Lipschitz continuous on all of $BV_{per}$ and the Lipschitz constant can be taken uniformly on any bounded subset.

Next we claim that
 that $S'(\eta) h (x)=  2 \pi \Xi(\eta) h(x)$. Fix $\eta(x),h(x) \in BV_{per}$. 
The fundamental theorem of calculus implies
\be\label{useful}
\Delta(x):=S(\eta+h)(x) - S(\eta)(x) - 2 \pi \Xi(\eta)h(x) = -4 \pi^2 \int_{0}^{h(x)} \int_0^s \sin(2 \pi  (\eta(x)+\tau) d\tau ds.
\ee
If we can show $\|\Delta\|_{BV_{per}} \le C(1 + \|\eta\|_{BV_{per}}) \| h \|_{BV_{per}}^2$ we will have established the claim.

Elementary estimates show 
$$
|\Delta(x)|=\left \vert S(\eta+h)(x) - S(\eta)(x) - 2 \pi \Xi(\eta)h(x)\right \vert \le C |h(x)|^2
$$
for some constant $C > 0$, independent of both $h$ and $\eta$. This implies
\be\label{L1 part}
\| \Delta \|_{L^1_{per}} = \|   S(\eta+h) - S(\eta) - 2 \pi \Xi(\eta)h\|_{L^1_{per}} \le C \| h^2 \|_{L^1_{per}} \le C \| h \|^2_{L^\infty_{per}} \le C \| h \|^2_{BV_{per}}.
\ee
Next we estimate
$$
V(\Delta) = \sup_{P \in \mathcal{P}} \sum_{P} \left \vert \Delta(x_i) - \Delta(x_{i-1}) \right \vert. 
$$
Then \eqref{useful} implies
$$
 \left \vert \Delta(x_i) - \Delta(x_{i-1}) \right \vert = 4 \pi^2 \left \vert 
 \int_{0}^{h(x_i)} \int_0^s \sin(2 \pi  (\eta(x_i)+\tau) d\tau ds -  \int_{0}^{h(x_{i-1})} \int_0^s \sin(2 \pi  (\eta(x_{i-1})+\tau) d\tau ds
 \right \vert 
$$
Adding zero, the triangle inequality and elementary properties of the integral, give:
\begin{multline}\label{thing}
 \left \vert \Delta(x_i) - \Delta(x_{i-1}) \right \vert = 4 \pi^2 \left \vert 
 \int_{h(x_{i-1})}^{h(x_i)} \int_0^s \sin(2 \pi  (\eta(x_i)+\tau) d\tau ds \right \vert \\ + 4 \pi^2  \left \vert  \int_{0}^{h(x_{i-1})} \int_0^s \left[\sin(2 \pi  (\eta(x_{i})+\tau)-\sin(2 \pi  (\eta(x_{i-1})+\tau)  \right]d\tau ds \right \vert. 
\end{multline}
Since $|\sin(\theta)| \le 1$ for any $\theta$ the first term is bounded by
$$
C \left \vert h(x_i) - h(x_{i-1})\right \vert \left \vert h(x_i) + h(x_{i-1}) \right \vert \le C \| h \|_{L^\infty_{per}} \left \vert h(x_i) - h(x_{i-1})\right \vert \le C \| h \|_{BV_{per}} \left \vert h(x_i) - h(x_{i-1})\right \vert.
$$
The constant $C>0$ is independent of $h$, $\eta$, $i$ or $P$.
And since $|\sin(\theta)-\sin(\phi)| \le  | \theta - \phi|$ for any $\theta,\phi$ the second term in \eqref{thing} is bounded by
$$
C \left\vert h(x_i)\right \vert ^2 \left \vert \eta(x_i) - \eta(x_{i-1}) \right \vert \le C\| h \|_{BV_{per}}^2 \left \vert \eta(x_i) - \eta(x_{i-1}) \right \vert
$$
The constant $C>0$ is independent of $h$, $\eta$, $i$ or $P$.

Thus we have
\begin{multline}
V(\Delta) \le C\| h \|_{BV_{per}} \sup_{P \in \mathcal{P}} \sum_{P} \left( 
|h(x_i) - h(x_{i-1})| + \| h \|_{BV_{per}} \left \vert \eta(x_i) - \eta(x_{i-1}) \right \vert
\right)\\ \le C\| h \|_{BV_{per}} V(h) +  C\| h \|^2_{BV_{per}} V(\eta) \le C(1 + \|\eta\|_{BV_{per}}) \| h \|^2_{BV_{per}}.
\end{multline}
Adding this to \eqref{L1 part} gives
$$
\| \Delta \|_{BV_{per}} \le  C(1 + \|\eta\|_{BV_{per}}) \| h \|^2_{BV_{per}}
$$
Thus we have, for any $\eta,\xi \in BV( \C)$
\be \label{S prime}
S'(\eta) \xi = 2 \pi \Xi(\eta) \xi
\ee
In an almost identical fashion we can show
\be\label{C prime}
\Xi'(\eta) \xi = -2 \pi S(\eta) \xi
\ee

The fact that $S$ and $\Xi$ are uniformly Lipschitz on any bounded subset of $BV_{per}$, together with \eqref{S prime} and \eqref{C prime}
and the fact that $BV_{per}$ is an algebra, is sufficient to conclude that $S'$ and $\Xi'$ are likewise uniformly Lipschitz on any bounded subset of $BV_{per}$.
Moreover,  \eqref{S prime} and \eqref{C prime} imply, for instance, that $S''(\eta)(\xi_1,\xi_2) = -4 \pi^2 S(\eta) \xi_1 \xi_2$, which, again,
is uniformly Lipschitz on bounded sets. This argument can extended to arbitrarily many derivatives of $S$ and $\Xi$ and thus completes the proof. \hfill $\square$

With Lemma \ref{substitution operators}, we can prove Lemma \ref{Phi}.

\noindent{\it Proof of Lemma \ref{Phi}:} 
First observe that the addition/subtraction of angles formulas for trigonometric functions give after recalling the defintions of $S$ and $\Xi$ above:
\begin{multline*}
\Phi(\eta)(x) = \int_I K(y-x) \sin(2\pi q(y-x)) \left[\Xi(\eta)(y) \Xi(\eta)(x) + S(\eta)(y) S(\eta)(x) \right] dy\\
+  \int_I K(y-x) \cos(2\pi q(y-x)) \left[S(\eta)(y) \Xi(\eta)(x) - S(\eta)(x) \Xi(\eta)(y) \right] dy
\end{multline*}
This can be written more compactly as:
$$
\Phi(\eta) = {1 \over 2 \pi} \Xi(\eta) [\tilde{K}^{(q)}* \Xi(\eta)]+ {1 \over 2 \pi} S(\eta) [\tilde{K}^{(q)}* S(\eta)]+ {1 \over 2 \pi} \Xi(\eta) [{K}^{(q)}*S(\eta)]- {1 \over 2 \pi} S(\eta) [{K}^{(q)}* \Xi(\eta)].
$$
where
$$
\tilde{K}^{(q)}(\theta) = -2\pi K(\theta) \sin(2\pi q \theta) \quad \text{and} \quad {K}^{(q)}(\theta) = 2 \pi K(\theta) \cos(2\pi q \theta). 
$$
Note that $\tilde{K}^{(q)}$ and $K^{(q)}$ are in $L^2(\C)$. The convolution estimates
$$
\| f * g \|_{BV_{per}} \le C \| f\|_{L^2(\C)} \|g \|_{BV_{per}} \quad \text{and} \quad \| f * g \|_{H^1_{per}} \le C \| f\|_{L^2(\C)} \|g \|_{H^1_{per}}
$$
imply that the linear maps $g \mapsto K^{(q)} * g$ and $g \mapsto \tilde{K}^{(q)} * g$ are bounded from $BV_{per}$ to itself and $H^1_{per}$ to itself.
Thus they are $C^\infty$ and all derivatives are globally Lipschitz. Since $\Xi$ and $S$ are $C^\infty$, this last result implies that
$\tilde{K}^{(q)}* \Xi(\eta)$, 
$\tilde{K}^{(q)}* S(\eta)$, 
${K}^{(q)}* \Xi(\eta)$, and
${K}^{(q)}*S(\eta)$ are $C^\infty$ as well, since each is a composition of smooth maps.

Finally the algebra estimates
$
\| f g \|_{BV_{per}} \le C \| f\|_{BV_{per}} \|g \|_{BV_{per}}
$
and
$
\| f g \|_{H^1_{per}} \le C \| f\|_{H^1_{per}} \|g \|_{H^1_{per}}
$
imply that the product of two smooth maps is smooth.
Thus we see that $\Phi(\eta)$ is $C^\infty$ since it is a linear combination of products and compositions of smooth maps.
All of its derivatives are uniformly Lipschitz
on any bounded subsets since its constituent parts are. \hfill $\square$

\section{Examples}\lbl{sec.examples}
\setcounter{equation}{0}

In this section, we apply Theorem~\ref{thm.stab} to establish stability of certain  
$q$--twisted states in the continuum limit of the KM on the nearest--neighbor 
and small--world graphs (cf. \eqref{Kur}, \eqref{Kr}), thus extending the linear
stability results obtained for these models in \cite{WilStr06, Med14c}. Throughout this
section, stability of twisted states is interpreted as stability with respect to perturbations
in $H^1_{per}$ and $BV_{per}$.

For convenience, we rewrite the continuum KM on small-world graphs
(cf. \eqref{Kur}, \eqref{Kr})
\be\lbl{KurNN}
\partial_t u(x,t) = \int_{I} K_{r,p}(y-x) \sin(2\pi(u(y,t)-u(x,t))) dy.
\ee
Note that for $p=0$ ($K_{r,0}(\cdot)=K_{r}(\cdot)$), \eqref{KurNN}
contains the KM on the nearest-neighbor graphs as a special case. 

The following theorem shows that $q$-twisted states are stable provided
$r>0$ and $p\ge 0$ are sufficiently small.

\begin{theorem}\lbl{thm.stabNN} Let $u^{(q)}$ be a $q$--twisted state solution
of \eqref{KurNN}.  
\begin{description}
\item[a)] $u^{(0)}$ is stable for any $r\in (0,1/2)$ and $p\in [0,1/2)$.
\item[b)] For $q\in \N$ there exist $r_q\in (0,1/2)$ and $p_q\in (0,1/2)$ such 
that $u^{(q)}$ is a stable steady-state solution of \eqref{KurNN} for any $r\in (0,r_q)$
and $p\in [0,p_q)$.
\end{description}
\end{theorem}

We precede the proof of Theorem~\ref{thm.stabNN} with the following auxiliary lemma,
whose proof is given at the end of this section.
\begin{lemma} \lbl{lem.sinc} 
Let 
$$
f(x):={\sin(x) \over x} \quad \text{and} \quad g(x,y) := f(y+x)-2f(y) +f(y-x) .
$$
There exist $X_1,Y_1 > 0$ and $\delta < 0$ such that
the following are true when $|y| \le Y_1$.
\begin{description}
\item[a)]  $g(x,y) = 0$ implies $x = 0$.
\item[b)]  $g(x,y) \le 0$ for all $x$.
\item[c)]  $g(x,y) \le \delta$ for $|x| \ge X_1$.
\end{description}
\end{lemma}

\pf (Theorem~\ref{thm.stabNN})
We prove the statements of the theorem for $p=0$  in part \textbf{A)}
below and then in part \textbf{B)} to show that these results continue to 
hold for small positive $p$.
\begin{description}
\item[A)] Suppose  $p=0$.
First, we specialize the stability condition \eqref{linear} for the KM \eqref{KurNN}. Using
\eqref{Kr}, we find
\be\lbl{Kr-tilde}
\tilde K_r(m)=\int_{-r}^r \cos(2\pi m s)ds = 2r f(2\pi m r), \; m\in \Z,
\ee
and rewrite \eqref{linear} for the model at hand
\be\lbl{linear-NN}
\lambda(q,m)= 2r g(m \rho,  q \rho) \le \lambda_q<0, \forall m\in \N, 
\ee
where $\rho:=2\pi r.$

We consider $q=0$ first. Let $0<\rho <\pi$ be arbitrary but fixed. In this case, 
$$
g(m\rho,0)=2\left( f(m\rho)-1\right).
$$
By Lemma~\ref{lem.sinc}\textbf{c},
$$ 
g(m\rho,0) \le \delta <0, \;\mbox{for}\; m\ge m_1:=\lceil X_1/\rho\rceil.
$$ 
Thus, \eqref{linear-NN} holds with 
$$
\lambda_0=(2r)^{-1}\max\left\{ \delta, \max_{1\le i\le m_1} 2\left( f(m\rho)-1\right)\right\} <0.
$$

Next, suppose $q\in \N$ is fixed. 
The quantity we are interested in
is
$$
\sup_{m \in \N} g(m\rho,q\rho).
$$
Let $\rho_q:=2Y_1/ q$ and
$$
\delta_q:=\max_{|x| \in  [\rho_q,X_1] } g(x,Y_1). 
$$
The first two conditions from the Lemma~\ref{lem.sinc} imply that $\delta_q < 0$, 
since we have $\rho_q>0$ and $g(x,Y_1)$ is a continuous function of $x$.
Let $\lambda_q:=\max\{ \delta_q, \delta \}/2r <0$.
Then we have
\be\lbl{sup-m}
\lambda(q,m)\le \sup_{m \ge 1} 2r g(m\rho, q\rho) \le \sup_{x \ge \rho_q} 2r g(x,Y_1)\le \lambda_q < 0.
\ee
Thus, \eqref{linear-NN} holds for all $0<\rho<\rho_q$.
\item[B)] Let $0<p<1/2$ and $0<r<1/2$. For
$$
K_{r,p}(x) = (1-2p)  \one_{(-r,r)}(x) +p \one_{(-1/2,1/2)}(x)
$$
we compute
$$
\tilde K_{r,p}(m)=(1-2p)\tilde K_r(m) +p \delta_{m,0},
$$
where $\one_A$ is the characteristic function of $A$, $\delta_{m,n}$ is the Kronecker delta,
and $\tilde K_r$ is given in \eqref{Kr-tilde}. Thus, \eqref{linear} becomes
\be\lbl{linear-SW}
\lambda(q,m;p)=(1-2p)\lambda(q,m)+p \delta_{q,0},
\ee
where $\lambda(q,m)$ is given in \eqref{linear-NN}.
Stability of the $0$-twisted state follows from the analysis in part \textbf{A)}.
For $q>0$, using \eqref{sup-m}, we conclude that
$$
\lambda(q,m;p)\le (1-2p)\lambda_q+p\delta_{m,q} \le \lambda_q/2\;\; \forall\;\; m\in\N,
$$
provided $0< p<-\lambda_q/(2(1-2\lambda_q))$.
\end{description}
\qed

\noindent \pf (Proof~Lemma~\ref{lem.sinc})
Note that $g(x,y)$ is even in both $x$ and $y$ so we only need to consider $x,y>0$.

Let $Y_{11} = \pi /2$. Since $f(y)$ is strictly decreasing on the interval $(0,\pi)$ we know that
$f(y) \ge  f(\pi/2) = 2/\pi$ when $|y| \le Y_{11}$. Thus $|y| \le Y_{11}$ implies 
$$
g(x,y) \le f(y+x) + f(y-x) -4/\pi \quad \text{for all $x$}.
$$

Next let $X_{1} := 5\pi/2$. If $0<y<Y_{11}=\pi/2$ then clearly $x>X_{1}$ gives $x > 2y$. Which means that
$$
x+y > x-y >x/2.
$$
We know that
$$
|f(x)| \le 1/|x|
\quad \text{ for all } 
x \ne 0.$$ Thus
\begin{eqnarray*}
|f(x+y)| & \le&  1/|x+y| \le 2/|x| \le 2/X_{1} = 4/5\pi \\
 |f(y-x)| &\le &1/|x-y| \le 2/|x| \le 2/X_{1} = 4/5\pi. 
\end{eqnarray*}
Thus we have
\be\label{conc 3}
|x| \ge X_{1} \text{ and } |y| \le Y_{11} \implies g(x,y) \le 8/5\pi - 4/\pi = -12/5\pi < 0.
\ee

Next we compute the Taylor series for $g(x,y)$ at $(0,0)$. It is:
$$
g(x,y)=x^2 \left(-{1 \over 3} + {1 \over 60} x^2 + {1 \over 10} y^2 \right) + O(|x,y|^4).
$$
Clearly $-{1 \over 3} + {1 \over 60} x^2 + {1 \over 10} y^2 < 0$ for $\|(x,y)\|$ sufficiently small. Thus we can conclude
that there exists $\rho_1 >0$ such that
\be\label{conc 1}
g(x,y) \le 0 \text{ for $\|(x,y)\|\le \rho_1$ with equality only if $x = 0$}.
\ee

Next note that $g(x,0) = 2f(x) - 2f(0) = 2 f(x) - 2$. The fact that $|f(x)| < 1$ for all $x\ne0$ means that $g(x,0) < 0$ for all $x \ne 0$.
Which means that there exists $\mu < 0$ such that $\sup_{x \in [\rho_1,X_1]} g(x,0) = \mu$. Since $g(x,y)$ is a smooth function 
and $[\rho_1,X_1]$ is compact in $\R^2$, we can conclude that there is open neighborhood of $[\rho_1,X_1]$ where $g(x,y)$ remains strictly negative.
That it to say there exists $\rho_2 > 0$ such that
\be\label{conc 2}
\sup_{ x \in [\rho_1 -\rho_2,X_1+\rho_2],|y| \le \rho_2} g(x,y) \le \mu/2<0.
\ee

Now put $Y_1:=\min\left\{Y_{11},\rho_1,\rho_2 \right\}$.  If $|y| \le Y_1$ and $|x|\ge X_1$ then \eqref{conc 3} gives us conclusion (c). If $|y| \le Y_1$ and $|x|\le X_1$ then
\eqref{conc 1} and \eqref{conc 2} give us conclusions (a) and (b).\\
\qed

\section{Conclusion}
\setcounter{section}{0}

Stability of a family of steady state solutions, such as twisted
states of the nonlocal equation \eqref{KurS}, provides valuable insights into the 
structure of the phase space and can be used for studying more complex 
dynamical regimes. Analysis of twisted states in the Kuramoto model of coupled 
phase oscillators has helped to understand better the link between the structure 
and dynamics in complex networks of interacting dynamical systems. 
It reveals a subtle relation between the fine properties
of the network topology and stability of steady state solutions in coupled systems
of nonlinear differential equations \cite{WilStr06, MedTan15b, Med14c}.

For the continuum model \eqref{KurS}, which approximates the dynamics of the
Kuramoto model on large graphs \cite{Med14a, Med14b}, Wiley, Strogatz, and 
Girvan found an elegant condition determining the stability of the twisted states
in this model \cite{WilStr06}. The condition relies on a set of linear inequalities 
in terms of the Fourier coefficients of the kernel of the integral operator in 
\eqref{KurS}. It applies to the Kuramoto model on a variety of Cayley and random
graphs \cite{MedTan15b}, including the small-world graphs \cite{Med14c}.
It is well-known that in infinite-dimmensional systems like \eqref{KurS}, linear 
stability does not automatically imply the nonlinear stability. In this work, we studied
under what conditions and in what sense twisted states are stable if the linear 
stability condition from \cite{WilStr06} holds. We found that for the problem at 
hand the linear stability implies nonlinear stability with respect to perturbations
in $H^1_{per}$ and $BV_{per}$. The latter result is important, because solutions of the 
initial value problems for \eqref{KurS} in general  have poor spatial regularity
and, thus, it is natural to consider stability with respect to rough perturbations.
The results of this work complement the linear stability analysis in \cite{WilStr06}
and provide a method for studying stability of spatial patterns in the continuum
Kuramoto system and related model.

\vskip 0.2cm
\noindent
{\bf Acknowledgements.} 
This work was supported in part by the NSF grants DMS-1105635 and DMS-1511488 (JDW)
and  DMS-1412066 (GM).

\bibliographystyle{amsplain}

\providecommand{\bysame}{\leavevmode\hbox to3em{\hrulefill}\thinspace}
\providecommand{\MR}{\relax\ifhmode\unskip\space\fi MR }
\providecommand{\MRhref}[2]{%
  \href{http://www.ams.org/mathscinet-getitem?mr=#1}{#2}
}
\providecommand{\href}[2]{#2}

\end{document}